\documentclass[preprint,prd,nofootinbib,tightenlines,amsmath]{revtex4}
\usepackage{bbm}
\usepackage{amsfonts}
\usepackage{mathrsfs}
\usepackage{graphicx}% Include figure files
\usepackage{bm}% bold math

\begin{document}
\baselineskip=15pt \parskip=5pt

\vspace*{3em}

\title{The new interaction suggested by the anomalous $^8$Be transition sets a rigorous constraint on the mass range of dark matter}

\author{Lian-Bao Jia$^1$}
\email{jialb@mail.nankai.edu.cn}

\author{Xue-Qian Li$^2$}
\email{lixq@nankai.edu.cn}

\affiliation{1. School of Science, Southwest University of Science and Technology, Mianyang
621010, China \\
2. School of Physics, Nankai University, Tianjin
300071, China \\}

\begin{abstract}

The WIMPs are considered one of the favorable dark matter (DM) candidates, but as the upper bounds on the interactions between DM and standard model (SM) particles obtained by the upgraded facilities of DM direct detections get lower and lower. Researchers turn their attentions to search for less massive DM candidates, i.e. light dark matter of MeV scale. The recently measured anomalous transition in $^8$Be suggests that there exists a vectorial boson which may mediate the interaction between DM and SM particles. Based on this scenario, we combine the relevant cosmological data to constrain the mass range of DM, and have found that there exists a model parameter space where the requirements are satisfied, a range of $10.4 \lesssim m_{\phi} \lesssim $ 16.7 MeV for scalar DM, and $13.6 \lesssim m_{V} \lesssim $ 16.7 MeV for vectorial DM is demanded. Then a possibility of directly detecting such light DM particles via the DM-electron scattering is briefly studied in this framework.

\end{abstract}

\maketitle

\section{Introduction}

For the time being, we still do not have solid knowledge on dark matter (DM). One of the preferable DM candidates is the weakly interacting massive particles (WIMPs), with WIMP masses of GeV-TeV scale. The recent DM direct detection experiments \cite{Angloher:2015ewa,Agnese:2015nto,Akerib:2015rjg,Aprile:2015uzo,Tan:2016zwf} set stringent constraints on the cross section of DM-target nucleus scattering for GeV-TeV scale DM, and the upper bound of the detection cross section will be reduced to the neutrino limit in next decade(s). On one aspect, the existence of DM is convinced by the astronomical observation, while on another aspect, the DM particles have not been detected by all the sophisticated experiments. One may ask if our conjecture on the potential mass range of DM is astray, which results in DM evading the present DM direct detections, namely, can the DM particles are much less massive to be in a sub-GeV range, e.g. in MeV (see Refs. \cite{Fayet:1980rr,Boehm:2002yz,Boehm:2003bt,Hooper:2003sh,Boehm:2003hm,Fayet:2004bw,Serpico:2004nm,Fayet:2006sp} for some earlier work). In this scenario, the interactions of the light DM particles just render the nucleus small recoil energies, which are not observable in available experiments for DM direct detections. In this work, we focus on the MeV scale light DM.

The issue concerning DM refers two aspects, one is the identities of DM, i.e what is (are) DM, and another aspect is how DM particles interact among themselves and with SM particles.
It is generally believed, the interactions related to the DM sector must be a new type (new types) beyond the standard model (BSM). In this work, to answer the first question, we do not priori assume
its identity, but let experimental data determine; to the second question, we look for a new BSM interaction which may offer an interpretation for the present observation.
The recent $^8$Be experiment has revealed at 6.8$\sigma$ an anomalous transition between an excited state $^8$Be$^\ast$ and the ground state $^8$Be \cite{Krasznahorkay:2015iga}. The authors  \cite{Krasznahorkay:2015iga,Feng:2016jff} argued that this anomaly may be due to the unknown nuclear reactions, but a more preferable possibility is that it is caused by emitting a vectorial boson $X$ during $^8Be^* \rightarrow ^8Be+ X$, which instantly decays into $e^+ e^-$ pair. The new boson $X$ may be the mediator that we look forward to between DM and SM particle interactions, and this probable is investigated in this paper. A fitted value of $X$ mass is $16.70 \pm 0.35 (stat)\pm 0.5 (sys)$ MeV \cite{Krasznahorkay:2015iga}, and in this work we adopt the central mass $m_X^{} \simeq 16.7$ MeV in calculations. The interactions of the vector boson $X$ with quarks and leptons via a scheme of BSM has been argued in the literatures \cite{Feng:2016jff,Gu:2016ege,Feng:2016ysn}. In this work, the vector boson $X$ discussed in Ref. \cite{Feng:2016jff} is of our concern.

For the scattering between possible scalar, vectorial, fermionic DM  and target nucleus, the spin-independent interaction induced by exchanging the vector boson $X$ is dominant (see e.g. Ref. \cite{Freytsis:2010ne}). The vector boson $X$ couples to electron and u,d quarks, and $X$ may also couples to the second and/or the third generation SM charged leptons and up type/down type quarks with equal couplings to the same type fermions (see, e.g. Ref. \cite{Gu:2016ege} for more discussions). For the thermally freeze-out DM with such couplings, the DM mass as low as 0.5 GeV has been excluded by the CRESST-II experiment \cite{Angloher:2015ewa}. Thus, the $X$-mediated sub-GeV DM needs more attention.

Here we focus on MeV scale DM. The energy released by DM annihilation can modify the cosmic microwave background (CMB), and the recent CMB measurement by the Planck satellite \cite{Ade:2015xua} sets a stringent bound on the s-wave annihilation of MeV-scale DM \cite{Ade:2015xua,Slatyer:2015jla}. For MeV DM with vector form interaction induced by $X$, the annihilation of fermionic DM pair is s-wave dominant, so is inconsistent with the CMB observation. Thus, the possibility of DM being fermions is disfavored. By contrast, p-wave annihilations of scalar and/or vector DM candidates at freeze out are tolerant by the CMB result. Thus, we concentrate on the case of scalar and vector DM, then the corresponding model parameter space will be derived.

For DM mass in the range of a few MeV/teens MeV, the big bang nucleosynthesis (BBN) and the effective number of relativistic neutrino $N_{eff}$ at recombination may be altered by the energy release from dark sector annihilations. Thus corresponding observation results will be taken into account to set a lower bound on DM mass.

As recoils of target nucleus are small, the scattering between DM and nucleus is not sensitive for DM in MeV region, thus the direct detection for DM would turn to the DM-electron scattering which might be employed for the light DM hunting, and the issue was investigated in Refs. \cite{Bernabei:2007gr,Dedes:2009bk,Kopp:2009et}. In this work the search for DM via its scattering with electron will be discussed for our concerned model.

This work is organized as follows. After this introduction, we present the concrete forms of interactions between SM and DM with new boson $X$ exchanged, and estimate the DM p-wave annihilation rate. Next we take into account the constraints by the BBN and CMB to set the mass range of DM, and numerically evaluate the DM-$X$ coupling for the DM mass range of concern. Then we analyze the detection possibility of the MeV DM via the DM-electron scattering. The last section is devoted to a brief conclusion and discussion.

\section{Interactions between SM and DM}

Based on the model where the new vector boson $X$ mediates interaction between the SM particles and scalar/vectorial DM, we will analyze the relevant issues. The couplings of $X$ with SM particles has been discussed in Ref. \cite{Feng:2016jff}. The effective $X$-DM coupling can be set in terms of the DM annihilation cross section at DM thermally freeze out.

\subsection{The couplings}

We suppose that $X$ mediates a BSM interaction where the new charge in $DM-X$ interaction is $e_D^{}$. The SM fermions are of equipped with also a new charge to couple to $X$ which is parameterized as $e \varepsilon_f$ (in unit of $e$), and $\varepsilon_f$ is relevant to the concerned fermion flavor. Let us first formulate the scattering amplitude between scalar DM and SM particles  caused by the new interaction where $X$ stands as the mediator. The new effective interaction is in the form
\begin{eqnarray}
\mathcal {L}^i_S &=& - e_D^{} X_{\mu}J^{\mu}_{DM} + e_D^2  X_{\mu} X^{\mu} \phi^\ast \phi - e \varepsilon_f X_{\mu}J^{\mu}_{SM} \,,
\end{eqnarray}
where $\phi$ is the scalar DM field. $J^{\mu}_{DM}$, $J^{\mu}_{SM}$ are the currents of scalar DM, SM fermions, respectively, with
\begin{eqnarray}
    J^{\mu}_{DM} &=& i [ \phi^\ast ( \partial^{\mu} \phi ) - (\partial^{\mu} \phi^\ast ) \phi ] \,, \quad   \rm scalar \, DM \,, \\
    J^{\mu}_{SM} &=& \Sigma_f \bar{f} \gamma^{\mu} f  \,, \quad \quad \quad \quad \quad  \,  \rm SM \, fermions\, .
\end{eqnarray}
To explain the $^8$Be anomalous transition, the $\varepsilon_f$ of the first generation fermion is derived and its value was presented in Ref. \cite{Feng:2016jff} as
\begin{eqnarray} \label{sm-coupling}
    &&\varepsilon_u \approx \pm 3.7 \times 10^{-3} \,, \quad   \varepsilon_d \approx \mp 7.4 \times 10^{-3} \,, \nonumber \\
    &&2 \times 10^{-4} \lesssim |\varepsilon_e| \lesssim 1.4 \times 10^{-3}  \,, \, \, |\varepsilon_{\nu} \varepsilon_e| \lesssim 7 \times 10^{-5} \, .
\end{eqnarray}
Moreover, if the vector boson $X$ couples to the muon with $|\varepsilon_\mu| \approx |\varepsilon_e|$, the discrepancy between theory and experiment in muon $g-2$ can be moderated \cite{Feng:2016jff}.

\begin{figure}[!htbp]
\includegraphics[width=3.2in]{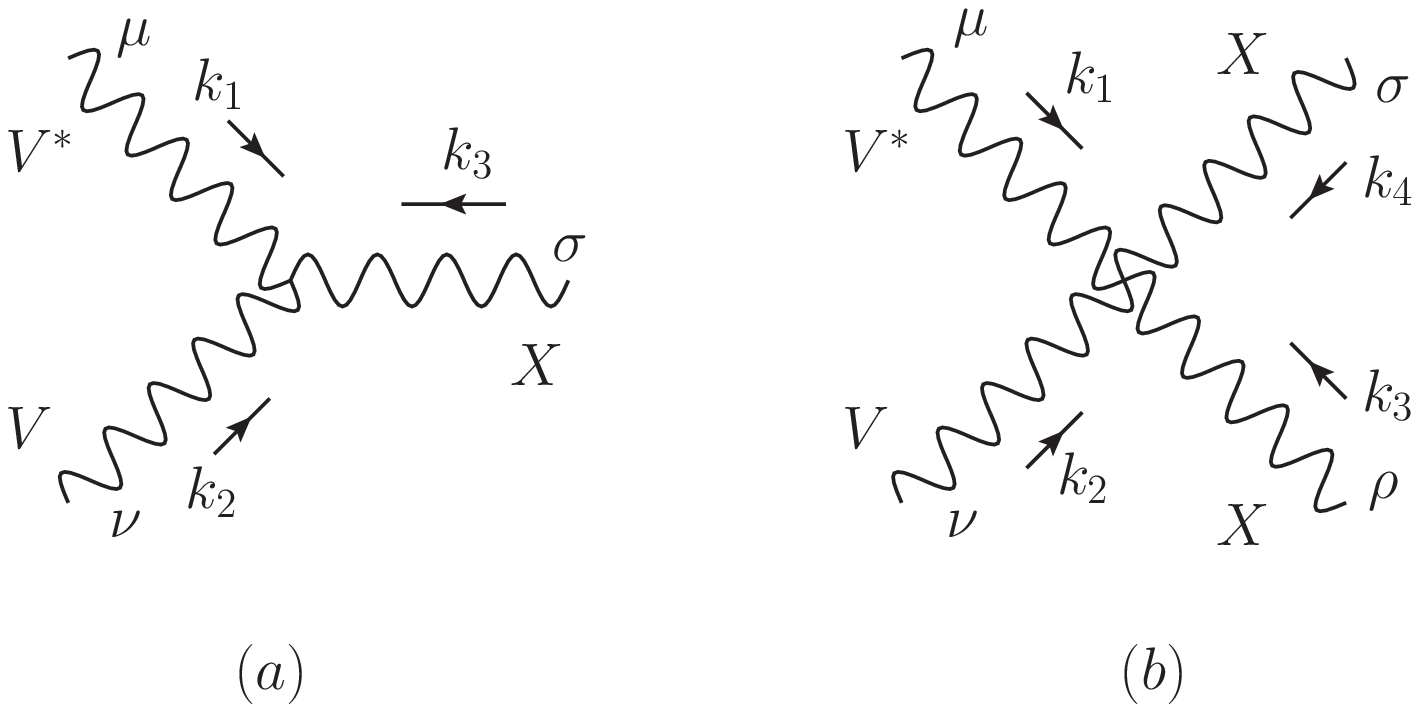} \vspace*{-1ex}
\caption{The vertexes of $V  V^\ast  X$, $V  V^\ast X X$.}\label{v-vertex}
\end{figure}

For the vectorial DM field $V$, the $V - X$ vertices are shown in Fig. \ref{v-vertex}. The $V  V^\ast X$ vertex is $- i e_D^{} [g^{\mu \nu}(k_2 - k_1)^\sigma + $ $g^{\nu \sigma}(k_3 - k_2)^\mu + g^{\sigma \mu}(k_1 - k_3)^\nu ]$, and the $V  V^\ast X X$ vertex is $ i e_D^{2} (g^{\mu \rho} g^{\nu \sigma} + $ $g^{\mu \sigma} g^{\nu \rho} - 2 g^{\mu \nu} g^{\rho \sigma} ) $. The couplings of $X$ in SM sector are the same as that of the scalar DM case.

\subsection{DM annihilations}

\begin{figure}[!htbp]
\includegraphics[width=3.6in]{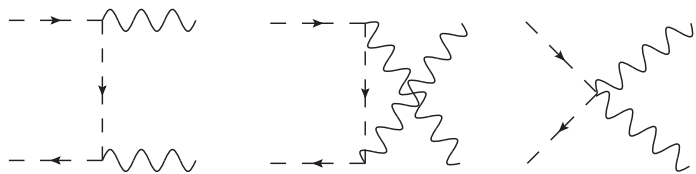} \vspace*{-1ex}
\caption{The annihilation $ \phi \phi^\ast \rightarrow X X$. The case of $ V V^\ast \rightarrow X X$ is similar.}\label{s-v-2}
\end{figure}

For scalar (vectorial) DM, the annihilation $ \phi \phi^\ast \rightarrow X \rightarrow f \bar{f}$ ($ V V^\ast \rightarrow X \to f \bar{f}$) is a p-wave process. When the scalar (vectorial) DM mass $m_\phi$ ($m_V$) is above the $X$ boson mass $m_X^{}$, the annihilation $ \phi \phi^\ast \rightarrow X X$ ($ V V^\ast \rightarrow X X$) portal is open, as shown in Fig. \ref{s-v-2}. However the analysis of Refs. \cite{Ade:2015xua,Slatyer:2015jla} indicate that the CMB measurement sets a stringent constraint on the MeV scale DM s-wave annihilation. For DM annihilation channels $e^+ e^-$ and $4 e$, the upper bounds from CMB on the s-wave annihilations of these two channels are as follows: e.g., for DM with the mass of 5 MeV, the cross sections are about below $ 2.7 \times 10^{-30}$, $ 4.3 \times 10^{-30}$ (cm$^3 / s$) for $e^+ e^-$, $4 e$, respectively; for DM with the mass of 500 MeV, the cross sections are about below $ 4.2 \times 10^{-28}$, $ 3.5 \times 10^{-28}$ (cm$^3 / s$) for $e^+ e^-$, $4 e$, respectively. For MeV scale DM, these constraints are much below the required thermally freeze-out annihilation cross section, and some tunings are needed if the DM s-wave annihilation exists. Thus for thermally freeze-out DM, to avoid the s-wave annihilation in the process $ \phi \phi^\ast \rightarrow X X$ ($ V V^\ast \rightarrow X X$), the constraint of $m_\phi$ ($m_V$) $ < m_X^{}$ is mandatory, i.e. the corresponding annihilation is kinematically closed. In addition, as indicated by the $^8$Be anomaly transition, the $X$ boson predominantly decays into $e^+ e^-$, and this implies that it cannot directly decay into DM, otherwise its decay procedure would be dominated by $X \to \phi \phi^\ast $ ($ V V^\ast $). Thus we must demand another constraint $m_\phi$ ($m_V$) $ > m_X^{}/2$. Therefore, a mass range of DM is $m_X^{}/2 < $ $m_\phi$ ($m_V$) $ < m_X^{}$, and the p-wave annihilation was overwhelming at DM freeze out.

\subsubsection{Scalar DM}

\begin{figure}[!htbp]
\includegraphics[width=3in]{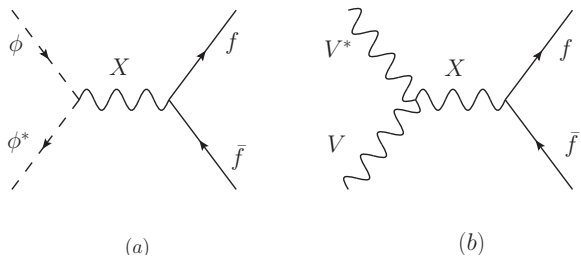} \vspace*{-1ex}
\caption{The annihilations of $ \phi \phi^\ast \rightarrow f \bar f$ (left) and $ V V^\ast  \rightarrow f \bar f$ (right).} \label{s-v-dm-ann}
\end{figure}

Let us first consider the scalar DM. In the mass range $m_X^{}/2 < $ $m_\phi$ $ < m_X^{}$, the s-channel annihilation $\phi \phi^\ast \rightarrow X\to f \bar{f}$ is overwhelming at DM freeze out, as shown in Fig. \ref{s-v-dm-ann} ($a$). In one initial DM particle rest frame, the scalar DM annihilation cross section can be written as
\begin{eqnarray}
\sigma_{ann} v_r = \frac{1}{2} \frac{e_D^2 e^2 \varepsilon_f^2}{( s - 2 m_\phi^2)}\frac{\beta_f }{8 \pi} \frac{(s-4m_\phi^2)[s - (s-4m_f^2)/3]}{(s-m_X^2)^2 + m_X^2 \Gamma_X^2}\, ,  \label{s-dm-ann}
\end{eqnarray}
where $v_r $ is the relative velocity of the two DM particles. The factor $\frac{1}{2}$ is due to the required $\phi \phi^\ast $ pair in annihilations, and $s $ is the total invariant squared mass. $\Gamma_X$ is the decay width of $X$, and $m_f$ is the mass of the final fermions. The phase space factor $\beta_f$ is
\begin{eqnarray}
\beta_f = \sqrt{1- \frac{4 m_f^2}{s} } \, .
\end{eqnarray}
Parameterizing Eq. (\ref{s-dm-ann}) in forms of
\begin{equation}\label{para}
\sigma_{ann} v_r =a + b v_r^2 + \mathcal {O} (v_r^4),
\end{equation}
with $s = 4 m_\phi^2 + m_\phi^2 v_r^2 + \mathcal {O} (v_r^4)$, we can obtain the result
\begin{eqnarray}
a = 0 \, , \quad     b =  \frac{e_D^2 e^2 \varepsilon_f^2 \beta_f }{8 \pi} \frac{ [m_\phi^2  - (m_\phi^2 -m_f^2)/3]}{(4 m_\phi^2 - m_X^2)^2 + m_X^2 \Gamma_X^2}\, .
\end{eqnarray}

With this parameterization, the thermally averaged annihilation cross section at temperature $T$ is \cite{Srednicki:1988ce,Gondolo:1990dk} $\langle \sigma_{ann} v_r \rangle \approx 6 b / x $,
with $x = m_\phi / T$. At DM thermally freeze-out temperature $T_f$, the parameter $x_f = m_\phi / T_f$ is \cite{Kolb:1990vq,Griest:1990kh}
\begin{eqnarray}
x_f \simeq \ln  0.038 c (c+2) \frac{g m_\phi m_{\rm {Pl}}  6 b / x_f }{\sqrt{g_\ast x_f}}\, ,
\end{eqnarray}
where $c$ is a parameter of $O(1)$, and we take $c = 1/2$ for numerical computations. $g$ is the degrees of freedom of DM, and $m_{\rm {Pl}} = 1.22 \times 10^{19}$ GeV is the Planck mass. $g_\ast$ is the total effective relativistic degrees of freedom at the temperature $T_f$, and we will adopt the data given by Ref. \cite{Drees:2015exa}. The relic density of DM is \cite{Kolb:1990vq,Griest:1990kh}
\begin{eqnarray}
\Omega_{DM} h^2 \simeq  \frac{1.07 \times 10^{9} x_f}{\sqrt{g_\ast} m_{\rm {Pl }} (GeV) (3 b / x_f) }  \, ,
\end{eqnarray}
where $h$ is the Hubble parameter (in units of 100 km/(s$\cdot$Mpc)).

\subsubsection{Vectorial DM}

Now consider the vectorial DM. In the mass range $m_X^{}/2 < $ $m_V$ $ < m_X^{}$, the annihilation $V V^\ast \rightarrow f \bar{f}$ is overwhelming at DM freeze out, as shown in Fig. \ref{s-v-dm-ann} ($b$). In one initial particle rest frame, the vectorial DM annihilation cross section is
\begin{eqnarray}
\sigma_{ann} v_r = \frac{1}{2} \frac{e_D^2 e^2 \varepsilon_f^2}{( s - 2 m_V^2)}\frac{\beta_f }{144 \pi} \frac{(s-4m_V^2)(s+2m_f^2)}{(s-m_X^2)^2 + m_X^2 \Gamma_X^2}[4 + \frac{7 s}{m_V^2} +  \frac{ s^2}{6 m_V^4}]\, .  \label{v-dm-ann}
\end{eqnarray}
Again parameterizing Eq. (\ref{v-dm-ann}) in forms of $\sigma_{ann} v_r =$ $a + b v_r^2 + \mathcal {O} (v_r^4)$, with $s = 4 m_V^2 + m_V^2 v_r^2 + \mathcal {O} (v_r^4)$, we have
\begin{eqnarray}
a = 0 \, , \quad     b =  \frac{e_D^2 e^2 \varepsilon_f^2 }{108 \pi} \frac{13 \beta_f (2 m_V^2 + m_f^2)}{(4 m_V^2 - m_X^2)^2 + m_X^2 \Gamma_X^2}\, .
\end{eqnarray}
The thermally averaged annihilation cross, the relic density of vectorial DM are similar to that we derived for scalar DM, replacing by corresponding input parameters.

\section{Analysis on $X$-DM coupling}

The energy released  from thermal MeV DM annihilation in the early universe can alter the BBN result and the effective number of relativistic neutrino $N_{eff}$. Even though the effects are not violent, it still can be employed to constrain the lower bound of DM mass. After the DM mass range being set, we will calculate the $X$-DM coupling by means of the DM thermally freeze-out annihilation cross section.

\subsection{DM mass with constraints of $N_{eff}$}

In the case of $m_X^{}/2 < $ $m_\phi$ ($m_V$) $ < m_X^{}$, the main annihilation product of DM is $e^+ e^-$. The DM annihilation might heat the electron-photon plasma before freeze out in the early universe. If this happens at the time that the neutrino decoupled from the hot bath, the ratio of the neutrino temperature relative to the photon temperature will be lowered, which causes a reduction of the number of the effective neutrino degrees of freedom \cite{Kolb:1986nf,Serpico:2004nm}. The abundances of light elements stemmed from the primordial nucleosynthesis and the CMB power spectra at the recombination epoch would also be affected. For electron neutrinos, a typical decoupling temperature is $T_d \sim$ 2.3 MeV \cite{Enqvist:1991gx}. The value $x_f$ of the thermally freeze-out DM is $x_f \sim$ 20. Thus, for the DM of concern, the freeze out of DM is supposed to be after neutrino decoupling, so the effects of DM annihilation need to be taken into account. For the new boson $X$, the decay width is
\begin{eqnarray}
 \Gamma_X \simeq \frac{e^2 \varepsilon_e^2 (m_X^2 + 2 m_e^2)}{12 \pi m_X^{}} \sqrt{1- \frac{4 m_e^2}{m_X^2}}.
\end{eqnarray}
With the mass $m_X^{} \gg T_d$ and $X$'s lifetime much less than 1 second, the contribution from $X$'s entropy to the BBN is negligible.

Here we focus on the constraints from the primordial abundances of light elements $^4$He and deuterium, denoted by $Y_p$ and $y_{DP}^{}$, respectively. The abundance values of $^4$He and deuterium are related to the baryon density $\omega_b \equiv \Omega_b h^2$ and the effective number of relativistic neutrinos $N_{eff}$ (or, in the form of the difference of $\Delta N_{eff} \equiv N_{eff} - 3.046$, where  $N_{eff} =$ 3.046 is the standard cosmological prediction value \cite{Dolgov:2002wy,Mangano:2005cc}). The abundances predicted by the BBN are parameterized as $Y_p$ ($\omega_b$, $\Delta N_{eff}$), $y_{DP}^{}$ ($\omega_b$, $\Delta N_{eff}$), and the corresponding Taylor expansion forms can be obtained with the PArthENoPE code \cite{Pisanti:2007hk}. If the value $\omega_b = 0.02226^{+0.00040}_{-0.00039}$ is adopted with the bounds of $Planck$ TT+lowP+BAO \cite{Ade:2015xua}, the value of $N_{eff}$ is also determined by the constraints of $^4$He and deuterium abundances. The range of $N_{eff}$ can be derived with the $Planck$ data, and that is \cite{Ade:2015xua}
\begin{eqnarray}
N_{eff} = \bigg \{ \begin{array}{cc}
  3.14^{+0.44}_{-0.43}\, \, \, & \rm{He}+\emph{Planck} \, \rm{TT+lowP+BAO} \,, \\
  3.01^{+0.38}_{-0.37}\, \, \, & \rm{D }+\emph{Planck} \, \rm{TT+lowP+BAO} \,, \label{He-D}
\end{array}
\end{eqnarray}
where the helium, deuterium abundances given by Aver et al. \cite{Aver:2013wba}, Cooke et al. \cite{Cooke:2013cba} are taken. The updated $Planck$-only constraint on $N_{eff}$ is \cite{Ade:2015xua}
\begin{eqnarray}
N_{eff} = 3.15 {\pm0.23}\, \, \, & Planck \, \rm{TT+lowP+BAO} \,. \label{Planck-only}
\end{eqnarray}
Considering Eqs. (\ref{He-D}), (\ref{Planck-only}), an lower bound $N_{eff} \gtrsim 2.9$ is taken in calculations.

\begin{figure}[!htbp]
\includegraphics[width=2.8in]{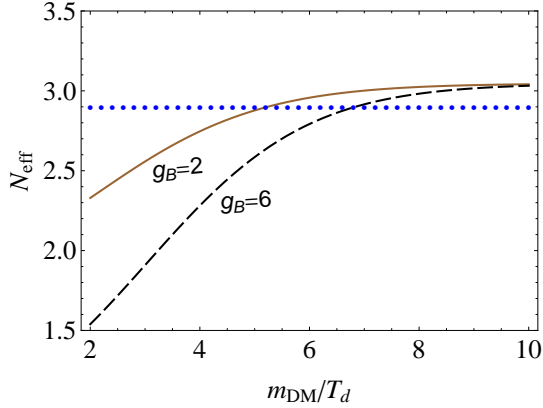} \vspace*{-1ex}
\caption{The effective number $ N_{eff}$ as a function of $m_{DM} / T_d$. The solid, dashed curves are the scalar, vectorial DM of concern, respectively. The dotted curve is for the lower bound $N_{eff} = 2.9$.} \label{neff-v}
\end{figure}

In the case that DM mainly couples to electron-photon plasma and DM particles freeze out later than the neutrino decoupling, the effective number $N_{eff}$ can be written as \cite{Ho:2012ug,Ho:2012br}
\begin{eqnarray}
N_{eff} = 3.046 \, [ \frac{I(0)}{I(T_d)}]^{\frac{4}{3}} \, \,,
\end{eqnarray}
where $I(T_\gamma)$ is given by
\begin{eqnarray}
I(T_\gamma) &=&  \frac{1}{T_\gamma^4} (\rho_{e^+ e^-} + \rho_{\gamma} + \rho_{DM}^{} + p_{e^+ e^-} + p_{\gamma} + p_{DM}^{})\, \,  \nonumber \\
&=& \frac{11}{45} \pi^2 + \frac{g}{2 \pi^2} \int^\infty_{y = 0} d y \frac{y^2}{e^{\, \xi} \pm 1} (\xi + \frac{y^2}{3 \xi}) \, ,
\end{eqnarray}
and
\begin{eqnarray}
 \xi = \sqrt{y^2 + (m_{DM}/{T_{\gamma}})^2} \, .
\end{eqnarray}
Here $T_\gamma$ is the photon temperature, and the integration variable is $y = p_{DM}/T_\gamma$. The plus/minus sign is for fermionic/bosonic DM particles, respectively. For bosonic DM of concern, the parameter values of the degrees of freedom $g_B^{} =$ 2, $g_B^{} =$ 6, the mass $m_{DM} =$ $m_\phi$, $m_V$ are corresponding to the scalar, vectorial DM, respectively. The effective number $N_{eff}$ as a function of $m_{DM} / T_d$ is shown in Fig. \ref{neff-v}. Taking the lower bound $N_{eff} \gtrsim 2.9$, we can obtain that $m_{DM} / T_d  \gtrsim$ 5.2, 6.8 for scalar, vectorial DM, respectively. As the neutrino decoupling is not a sudden process (for more details, see e.g. Refs. \cite{Enqvist:1991gx,Dolgov:2002wy,Mangano:2005cc,Hannestad:2001iy}), here we take $T_d  \gtrsim$ 2 MeV as a lower bound. Thus, the mass range of DM is derived,
\begin{eqnarray}
 \bigg \{ \begin{array}{cc}
  10.4 \lesssim m_{\phi} \lesssim 16.7 ~~ \rm(MeV) \,  &  \rm scalar \, DM \,, \\
  13.6 \lesssim m_{V}    \lesssim 16.7 ~~ \rm(MeV) \,  &  \rm vectorial \, DM \,.
\end{array}
\end{eqnarray}

\subsection{Numerical result for the $X$-DM coupling}

\begin{figure}[!htbp]
\includegraphics[width=3.2in]{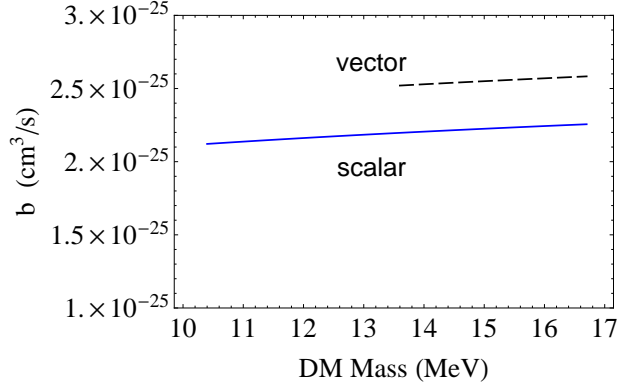} \vspace*{-1ex}
\caption{The parameter $b$ as a function of DM mass. The solid, dashed curves are the scalar, vectorial DM of concern, respectively.} \label{b-value}
\end{figure}

\begin{figure}[!htbp]
\includegraphics[width=2.8in]{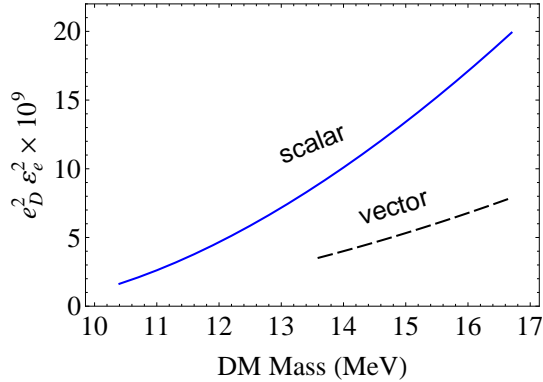} \vspace*{-1ex}
\caption{The values of $e_D^2  \varepsilon_e^2$ as a function of DM mass. The solid, dashed curves are the scalar, vectorial DM, respectively.} \label{num-coupling}
\end{figure}

As the DM mass range being set, we turn to investigate  the $X$-DM coupling. The DM relic density is $0.1197 \pm 0.0042$ \cite{Ade:2015xua}. According to the DM thermally averaged annihilation cross section $\langle \sigma_{ann} v_r \rangle \approx 6 b / x_f $ at $T_f$, the numerical results of $b$ are shown in Fig. \ref{b-value}, with the solid, dashed curves corresponding to the scalar, vectorial DM, respectively. After the values of $b$ defined in Eq. (\ref{para}) is obtained, and then the $X$-DM coupling couplings is also determined. The numerical results of $e_D^2 \varepsilon_e^2$ are depicted in Fig. \ref{num-coupling}. Considering the value of $\varepsilon_e$ given by Eq. (\ref{sm-coupling}), we can obtain
$e_D^2 / 4 \pi <$ 1, and thus the $X$-DM coupling is sufficiently small that the perturbation may apply.

\section{DM-electron scattering}

Now let us turn to investigate the possibility of detecting the light DM of MeV scale by the earth detector.

For the light DM particles, since the recoil of the target nucleus is too small to be substantially observed, one may not detect arrival of DM via the scattering between the MeV DM and target nucleus. Instead, the DM-electron scattering can be employed for the MeV DM hunting. The DM-electron scattering has been investigated in Refs. \cite{Bernabei:2007gr,Dedes:2009bk,Kopp:2009et}. The target atomic electron is in a bound state, and the typical momentum transfer $q$ is of order $\alpha m_e$
as a few eV, which may cause excitation/ionization of the electron in inelastic scattering processes. In this work, we study the signals of individual electrons induced by DM-electron scattering. Here, we take the form of the DM-electron scattering cross section as given by Ref. \cite{Essig:2011nj}, and for scalar DM, that is
\begin{eqnarray}
\bar{\sigma}_e &=& \frac{\mu_{\phi e}^2}{16 \pi m_{\phi}^2 m_{e}^2} \overline{|\mathcal{M}_{\phi e} (q)|^2}\big|_{q^2 = \alpha^2 m_e^2} \times |F_{DM}(q)|^2\, \\ \nonumber
    &\simeq& \frac{4 \alpha e_D^2  \varepsilon_e^2  \mu_{\phi e}^2}{m_X^4}   \, ,
\end{eqnarray}
with $\mu_{\phi e}$ being the $\phi$-electron reduced mass, and $F_{DM}(q) \simeq 1$ for $m_X^{} \gg \alpha m_e$.

For vectorial DM, the DM-electron scattering cross section is
\begin{eqnarray}
\bar{\sigma}_e &=& \frac{\mu_{V e}^2}{16 \pi m_{V}^2 m_{e}^2} \overline{|\mathcal{M}_{V e} (q)|^2}\big|_{q^2 = \alpha^2 m_e^2} \times |F_{DM}(q)|^2\, \\ \nonumber
    &\simeq& \frac{4 \alpha e_D^2  \varepsilon_e^2  \mu_{V e}^2}{m_X^4}   \, ,
\end{eqnarray}
with $\mu_{V e}^{}$ being the $V$-electron reduced mass.

\begin{figure}[!htbp]
\includegraphics[width=3.2in]{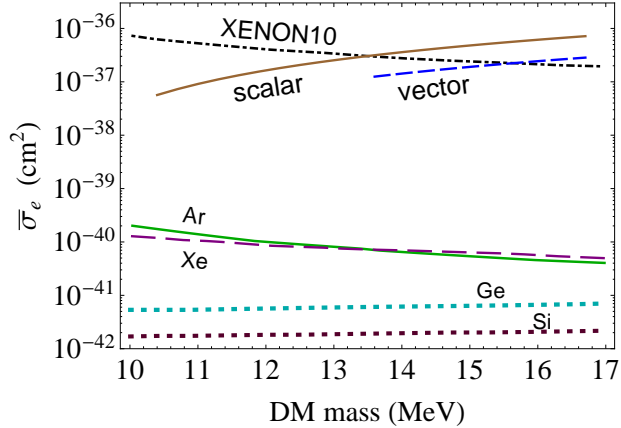} \vspace*{-1ex}
\caption{The DM-electron scattering cross section $\bar{\sigma}_e$ as a function of DM mass with the parameter $F_{DM}(q) = 1$. The upper solid, upper dashed curves are the scalar, vectorial DM, respectively. The dot-dashed curve is the excluded bound set by the XENON10 data \cite{Essig:2012yx}. The lower solid, lower dashed curves are the 95\% confidence level exclusion reach of single electron detections set by the 1 kg$\cdot$year exposure of Ar, Xe \cite{Essig:2011nj}, respectively. The upper, lower square curves are the 95\% confidence level exclusion reach of single electron detections set by the 1 kg$\cdot$year exposure of Ge, Si \cite{Essig:2015cda}, respectively.} \label{dm-ele-cs}
\end{figure}

As the value of $e_D^2  \varepsilon_e^2$ is fixed, the DM-electron scattering cross section $\bar{\sigma}_e$ can be obtained. The numerical result of $\bar{\sigma}_e$ is shown in Fig. \ref{dm-ele-cs}, where it is noted that the scattering cross section is independent of the momentum transfer ($F_{DM}(q) = 1$). The upper solid, upper dashed curves are for the scalar, vectorial DM, respectively, and the dot-dashed curve is the excluded bound set by the XENON10 data \cite{Essig:2012yx}. It can be seen that, considering the constraint of XENON10, there exists parameter spaces for scalar, vectorial DM to satisfy the constraints.

Now we give a brief discussion about the background in the DM-electron scattering. One irreducible background is from the neutrino-electron scattering, which sets the ultimate limit to the sub-GeV DM direct detections. Fortunately, the DM annual modulation effect from the motion of the earth can be employed to reduce the neutrino background \cite{Drukier:1986tm,Essig:2011nj,Lee:2015qva}. The teens MeV DM of concern could be probed via the inelastic processes of DM-electron scatterings, e.g. the individual electron signals by the future noble gas and semiconductor targets. For Ar, Xe \cite{Essig:2011nj} and Ge, Si \cite{Essig:2015cda} with 1 kg$\cdot$year exposure, the exclusion reach at 95\% confidence level via single electron detections are also shown in Fig. \ref{dm-ele-cs}. Further explorations of DM-electron scatterings are needed, both in theory and experiment.

\section{Conclusion and discussion}

The MeV scalar and vectorial DM has been studied in this work, with the new boson $X$ indicated by the $^8$Be anomalous transition being the mediator. Considering the constraints of the DM direct detection and CMB observation, we find that for the case of $m_X^{}/2 < $ $m_\phi$ ($m_V$) $ < m_X^{}$, the p-wave dominant annihilation of DM at freeze out does not conflict with the observed data so far. The primordial abundances of light elements and the effective number of relativistic neutrino $N_{eff}$ at recombination are sensitive to the DM with the mass of a few MeV to teens MeV, thus the corresponding observed results have been employed to set a lower bound on the DM mass. Taking the combined lower bounds $N_{eff} \gtrsim 2.9$ and the neutrino decoupling temperature $T_d  \gtrsim$ 2 MeV, we derive a mass range of DM: $10.4 \lesssim m_{\phi} \lesssim 16.7 $ MeV for scalar DM, and $13.6 \lesssim m_{V} \lesssim 16.7$ MeV for vectorial DM.

For the teens MeV scalar, vectorial DM of concern, the numerical result of the DM-$X$ coupling is derived in terms of the DM thermally averaged annihilation cross section. Once this coupling is set, the strength of the interaction between DM and SM particles is determined.

The DM-electron scattering is employed for the teens MeV DM hunting. We investigate on the signal of the individual electrons in DM-electron scattering, and the scattering cross section $\bar{\sigma}_e$ is calculated for the DM mass range of concern. We find that, considering the constraint of XENON10, there are still parameter spaces left for the teens MeV scalar, vectorial DM to be observed. Beside the individual electrons,  signals of individual photons, individual ions, and heat/phonons can also be employed to explore the MeV DM-electron scattering (see. e.g. Ref. \cite{Essig:2011nj,Derenzo:2016fse} for more), even though the ion signal is probably too weak for detection. The teens MeV DM of concern could be probed by the future noble gas and semiconductor targets via the DM-electron scattering. In fact, the wave function of electron in the bound state  for a certain target material needs to be considered to guarantee the prediction power. It is noted that the detection possibilities and efficiency of DM are target dependent.

As discussed in Ref. \cite{Chen:2016dhm}, the new boson $X$ may be detectable at the $e^+ e^-$ collider, such as BESIII and BaBar. The new boson $X$ may also give an interpretation about the NuTeV anomaly \cite{Liang:2016ffe}. For the teens MeV scalar, vectorial DM of concern, further investigation both in theory and experiment aspects are needed. We look forward to the exploration of the $X$-portal DM in the future.

\acknowledgments \vspace*{-3ex} This work was partially supported by the National Natural Science Foundation of China under Contract No. 11505144, 11375128 and 11135009, and the Research Fund for the Doctoral Program of the Southwest University of Science and Technology under Contract No. 15zx7102.


\begin{thebibliography}{0} \vspace*{-2ex}




%\cite{Angloher:2015ewa}
\bibitem{Angloher:2015ewa}
  G.~Angloher {\it et al.} [CRESST Collaboration],
  %``Results on light dark matter particles with a low-threshold CRESST-II detector,''
  Eur.\ Phys.\ J.\ C {\bf 76} (2016) no.1,  25
  [arXiv:1509.01515 [astro-ph.CO]].


%\cite{Agnese:2015nto}
\bibitem{Agnese:2015nto}
  R.~Agnese {\it et al.} [SuperCDMS Collaboration],
  %``New Results from the Search for Low-Mass Weakly Interacting Massive Particles with the CDMS Low Ionization Threshold Experiment,''
  Phys.\ Rev.\ Lett.\  {\bf 116} (2016) no.7,  071301
  [arXiv:1509.02448 [astro-ph.CO]].


%\cite{Akerib:2015rjg}
\bibitem{Akerib:2015rjg}
  D.~S.~Akerib {\it et al.} [LUX Collaboration],
  %``Improved Limits on Scattering of Weakly Interacting Massive Particles from Reanalysis of 2013 LUX Data,''
  Phys.\ Rev.\ Lett.\  {\bf 116} (2016) no.16,  161301
  [arXiv:1512.03506 [astro-ph.CO]].



%\cite{Aprile:2015uzo}
\bibitem{Aprile:2015uzo}
  E.~Aprile {\it et al.} [XENON Collaboration],
  %``Physics reach of the XENON1T dark matter experiment,''
  JCAP {\bf 1604}, no. 04, 027 (2016)
  [arXiv:1512.07501 [physics.ins-det]].


%\cite{Tan:2016zwf}
\bibitem{Tan:2016zwf}
  A.~Tan {\it et al.} [PandaX-II Collaboration],
  %``Dark Matter Results from First 98.7-day Data of PandaX-II Experiment,''
  Phys.\ Rev.\ Lett.\  {\bf 117} (2016) no.12,  121303
  [arXiv:1607.07400 [hep-ex]].


%\cite{Fayet:1980rr}
\bibitem{Fayet:1980rr}
  P.~Fayet,
  %``On the Search for a New Spin 1 Boson,''
  Nucl.\ Phys.\ B {\bf 187} (1981) 184.

%\cite{Boehm:2002yz}
\bibitem{Boehm:2002yz}
  C.~Boehm, T.~A.~Ensslin and J.~Silk,
  %``Can Annihilating dark matter be lighter than a few GeVs?,''
  J.\ Phys.\ G {\bf 30} (2004) 279
  [astro-ph/0208458].

%\cite{Boehm:2003bt}
\bibitem{Boehm:2003bt}
  C.~Boehm, D.~Hooper, J.~Silk, M.~Casse and J.~Paul,
  %``MeV dark matter: Has it been detected?,''
  Phys.\ Rev.\ Lett.\  {\bf 92} (2004) 101301
  [astro-ph/0309686].

%\cite{Hooper:2003sh}
\bibitem{Hooper:2003sh}
  D.~Hooper, F.~Ferrer, C.~Boehm, J.~Silk, J.~Paul, N.~W.~Evans and M.~Casse,
  %``Possible evidence for MeV dark matter in dwarf spheroidals,''
  Phys.\ Rev.\ Lett.\  {\bf 93} (2004) 161302
  [astro-ph/0311150].


%\cite{Boehm:2003hm}
\bibitem{Boehm:2003hm}
  C.~Boehm and P.~Fayet,
  %``Scalar dark matter candidates,''
  Nucl.\ Phys.\ B {\bf 683} (2004) 219
  [hep-ph/0305261].


%\cite{Fayet:2004bw}
\bibitem{Fayet:2004bw}
  P.~Fayet,
  %``Light spin 1/2 or spin 0 dark matter particles,''
  Phys.\ Rev.\ D {\bf 70} (2004) 023514
  [hep-ph/0403226].

%\cite{Serpico:2004nm}
\bibitem{Serpico:2004nm}
  P.~D.~Serpico and G.~G.~Raffelt,
  %``MeV-mass dark matter and primordial nucleosynthesis,''
  Phys.\ Rev.\ D {\bf 70} (2004) 043526
  [astro-ph/0403417].


%\cite{Fayet:2006sp}
\bibitem{Fayet:2006sp}
  P.~Fayet,
  %``Constraints on Light Dark Matter and U bosons, from psi, Upsilon, K+, pi0, eta and eta-prime decays,''
  Phys.\ Rev.\ D {\bf 74} (2006) 054034
  [hep-ph/0607318].



%\cite{Krasznahorkay:2015iga}
\bibitem{Krasznahorkay:2015iga}
  A.~J.~Krasznahorkay {\it et al.},
  %``Observation of Anomalous Internal Pair Creation in Be8 : A Possible Indication of a Light, Neutral Boson,''
  Phys.\ Rev.\ Lett.\  {\bf 116} (2016) no.4,  042501
  [arXiv:1504.01527 [nucl-ex]].


%\cite{Feng:2016jff}
\bibitem{Feng:2016jff}
  J.~L.~Feng, B.~Fornal, I.~Galon, S.~Gardner, J.~Smolinsky, T.~M.~P.~Tait and P.~Tanedo,
  %``Protophobic Fifth Force Interpretation of the Observed Anomaly in $^8$Be Nuclear Transitions,''
  Phys.\ Rev.\ Lett.\  {\bf 117} (2016) no.7,  071803
  [arXiv:1604.07411 [hep-ph]].


%\cite{Gu:2016ege}
\bibitem{Gu:2016ege}
  P.~H.~Gu and X.~G.~He,
  %``Realistic model for a fifth force explaining anomaly in ${^8Be^*} \to {^8Be} \;{e^+e^-}$ Decay,''
  arXiv:1606.05171 [hep-ph].


%\cite{Feng:2016ysn}
\bibitem{Feng:2016ysn}
  J.~L.~Feng, B.~Fornal, I.~Galon, S.~Gardner, J.~Smolinsky, T.~M.~P.~Tait and P.~Tanedo,
  %``Particle Physics Models for the 17 MeV Anomaly in Beryllium Nuclear Decays,''
  arXiv:1608.03591 [hep-ph].


%\cite{Freytsis:2010ne}
\bibitem{Freytsis:2010ne}
  M.~Freytsis and Z.~Ligeti,
  %``On dark matter models with uniquely spin-dependent detection possibilities,''
  Phys.\ Rev.\ D {\bf 83} (2011) 115009
  [arXiv:1012.5317 [hep-ph]].


%\cite{Ade:2015xua}
\bibitem{Ade:2015xua}
  P.~A.~R.~Ade {\it et al.} [Planck Collaboration],
  %``Planck 2015 results. XIII. Cosmological parameters,''
  arXiv:1502.01589 [astro-ph.CO].


%\cite{Slatyer:2015jla}
\bibitem{Slatyer:2015jla}
  T.~R.~Slatyer,
  %``Indirect dark matter signatures in the cosmic dark ages. I. Generalizing the bound on s-wave dark matter annihilation from Planck results,''
  Phys.\ Rev.\ D {\bf 93} (2016) no.2,  023527
  [arXiv:1506.03811 [hep-ph]].



%\cite{Bernabei:2007gr}
\bibitem{Bernabei:2007gr}
  R.~Bernabei {\it et al.},
  %``Investigating electron interacting dark matter,''
  Phys.\ Rev.\ D {\bf 77} (2008) 023506
  [arXiv:0712.0562 [astro-ph]].


%\cite{Dedes:2009bk}
\bibitem{Dedes:2009bk}
  A.~Dedes, I.~Giomataris, K.~Suxho and J.~D.~Vergados,
  %``Searching for Secluded Dark Matter via Direct Detection of Recoiling Nuclei as well as Low Energy Electrons,''
  Nucl.\ Phys.\ B {\bf 826} (2010) 148
  [arXiv:0907.0758 [hep-ph]].


%\cite{Kopp:2009et}
\bibitem{Kopp:2009et}
  J.~Kopp, V.~Niro, T.~Schwetz and J.~Zupan,
  %``DAMA/LIBRA and leptonically interacting Dark Matter,''
  Phys.\ Rev.\ D {\bf 80} (2009) 083502
  [arXiv:0907.3159 [hep-ph]].


%\cite{Srednicki:1988ce}
\bibitem{Srednicki:1988ce}
  M.~Srednicki, R.~Watkins and K.~A.~Olive,
  %``Calculations of Relic Densities in the Early Universe,''
  Nucl.\ Phys.\ B {\bf 310} (1988) 693.


%\cite{Gondolo:1990dk}
\bibitem{Gondolo:1990dk}
  P.~Gondolo and G.~Gelmini,
  %``Cosmic abundances of stable particles: Improved analysis,''
  Nucl.\ Phys.\ B {\bf 360} (1991) 145.


%\cite{Kolb:1990vq}
\bibitem{Kolb:1990vq}
  E.~W.~Kolb and M.~S.~Turner,
  %``The Early Universe,''
  Front.\ Phys.\  {\bf 69} (1990) 1.


%\cite{Griest:1990kh}
\bibitem{Griest:1990kh}
  K.~Griest and D.~Seckel,
  %``Three exceptions in the calculation of relic abundances,''
  Phys.\ Rev.\ D {\bf 43} (1991) 3191.


%\cite{Drees:2015exa}
\bibitem{Drees:2015exa}
  M.~Drees, F.~Hajkarim and E.~R.~Schmitz,
  %``The Effects of QCD Equation of State on the Relic Density of WIMP Dark Matter,''
  JCAP {\bf 1506} (2015) no.06,  025
  [arXiv:1503.03513 [hep-ph]].


%\cite{Kolb:1986nf}
\bibitem{Kolb:1986nf}
  E.~W.~Kolb, M.~S.~Turner and T.~P.~Walker,
  %``The Effect of Interacting Particles on Primordial Nucleosynthesis,''
  Phys.\ Rev.\ D {\bf 34} (1986) 2197.


%\cite{Enqvist:1991gx}
\bibitem{Enqvist:1991gx}
  K.~Enqvist, K.~Kainulainen and V.~Semikoz,
  %``Neutrino annihilation in hot plasma,''
  Nucl.\ Phys.\ B {\bf 374} (1992) 392.


%\cite{Dolgov:2002wy}
\bibitem{Dolgov:2002wy}
  A.~D.~Dolgov,
  %``Neutrinos in cosmology,''
  Phys.\ Rept.\  {\bf 370} (2002) 333
  [hep-ph/0202122].


%\cite{Mangano:2005cc}
\bibitem{Mangano:2005cc}
  G.~Mangano, G.~Miele, S.~Pastor, T.~Pinto, O.~Pisanti and P.~D.~Serpico,
  %``Relic neutrino decoupling including flavor oscillations,''
  Nucl.\ Phys.\ B {\bf 729} (2005) 221
  [hep-ph/0506164].


%\cite{Pisanti:2007hk}
\bibitem{Pisanti:2007hk}
  O.~Pisanti, A.~Cirillo, S.~Esposito, F.~Iocco, G.~Mangano, G.~Miele and P.~D.~Serpico,
  %``PArthENoPE: Public Algorithm Evaluating the Nucleosynthesis of Primordial Elements,''
  Comput.\ Phys.\ Commun.\  {\bf 178} (2008) 956
  [arXiv:0705.0290 [astro-ph]].



%\cite{Aver:2013wba}
\bibitem{Aver:2013wba}
  E.~Aver, K.~A.~Olive, R.~L.~Porter and E.~D.~Skillman,
  %``The primordial helium abundance from updated emissivities,''
  JCAP {\bf 1311} (2013) 017
  [arXiv:1309.0047 [astro-ph.CO]].


%\cite{Cooke:2013cba}
\bibitem{Cooke:2013cba}
  R.~Cooke, M.~Pettini, R.~A.~Jorgenson, M.~T.~Murphy and C.~C.~Steidel,
  %``Precision measures of the primordial abundance of deuterium,''
  Astrophys.\ J.\  {\bf 781} (2014) no.1,  31
  [arXiv:1308.3240 [astro-ph.CO]].


%\cite{Ho:2012ug}
\bibitem{Ho:2012ug}
  C.~M.~Ho and R.~J.~Scherrer,
  %``Limits on MeV Dark Matter from the Effective Number of Neutrinos,''
  Phys.\ Rev.\ D {\bf 87} (2013) no.2,  023505
  [arXiv:1208.4347 [astro-ph.CO]].

%\cite{Ho:2012br}
\bibitem{Ho:2012br}
  C.~M.~Ho and R.~J.~Scherrer,
  %``Sterile Neutrinos and Light Dark Matter Save Each Other,''
  Phys.\ Rev.\ D {\bf 87} (2013) no.6,  065016
  [arXiv:1212.1689 [hep-ph]].


%\cite{Hannestad:2001iy}
\bibitem{Hannestad:2001iy}
  S.~Hannestad,
  %``Oscillation effects on neutrino decoupling in the early universe,''
  Phys.\ Rev.\ D {\bf 65} (2002) 083006
  [astro-ph/0111423].



%\cite{Essig:2011nj}
\bibitem{Essig:2011nj}
  R.~Essig, J.~Mardon and T.~Volansky,
  %``Direct Detection of Sub-GeV Dark Matter,''
  Phys.\ Rev.\ D {\bf 85} (2012) 076007
  [arXiv:1108.5383 [hep-ph]].



%\cite{Essig:2012yx}
\bibitem{Essig:2012yx}
  R.~Essig, A.~Manalaysay, J.~Mardon, P.~Sorensen and T.~Volansky,
  %``First Direct Detection Limits on sub-GeV Dark Matter from XENON10,''
  Phys.\ Rev.\ Lett.\  {\bf 109} (2012) 021301
  [arXiv:1206.2644 [astro-ph.CO]].


%\cite{Lee:2015qva}
\bibitem{Lee:2015qva}
  S.~K.~Lee, M.~Lisanti, S.~Mishra-Sharma and B.~R.~Safdi,
  %``Modulation Effects in Dark Matter-Electron Scattering Experiments,''
  Phys.\ Rev.\ D {\bf 92} (2015) no.8,  083517
  [arXiv:1508.07361 [hep-ph]].



%\cite{Drukier:1986tm}
\bibitem{Drukier:1986tm}
  A.~K.~Drukier, K.~Freese and D.~N.~Spergel,
  %``Detecting Cold Dark Matter Candidates,''
  Phys.\ Rev.\ D {\bf 33} (1986) 3495.


%\cite{Essig:2015cda}
\bibitem{Essig:2015cda}
  R.~Essig, M.~Fernandez-Serra, J.~Mardon, A.~Soto, T.~Volansky and T.~T.~Yu,
  %``Direct Detection of sub-GeV Dark Matter with Semiconductor Targets,''
  JHEP {\bf 1605} (2016) 046
  [arXiv:1509.01598 [hep-ph]].


%\cite{Derenzo:2016fse}
\bibitem{Derenzo:2016fse}
  S.~Derenzo, R.~Essig, A.~Massari, A.~Soto and T.~T.~Yu,
  %``Direct Detection of sub-GeV Dark Matter with Scintillating Targets,''
  arXiv:1607.01009 [hep-ph].


%\cite{Chen:2016dhm}
\bibitem{Chen:2016dhm}
  L.~B.~Chen, Y.~Liang and C.~F.~Qiao,
  %``X(16.7) Production in Electron-Positron Collision,''
  arXiv:1607.03970 [hep-ph].


%\cite{Liang:2016ffe}
\bibitem{Liang:2016ffe}
  Y.~Liang, L.~B.~Chen and C.~F.~Qiao,
  %``X(16.7) as the Solution of NuTeV Anomaly,''
  arXiv:1607.08309 [hep-ph].
























%%%%%%%%%%%%%%%%%%%%%%%%%%%%%%%%










% New references












\end{thebibliography}
\end{document}